
\font\titlefont = cmr10 scaled\magstep 4
 2
\font\sectionfont = cmr10
\font\littlefont = cmr5 
\font\eightrm = cmr8

\newcount\tcflag
\tcflag = 0  

\ifnum\tcflag = 0 \magnification = 1200 \fi  

\global\baselineskip = 1.2\baselineskip 
\global\parskip = 4pt plus 0.3pt 
\global\abovedisplayskip = 18pt plus3pt minus9pt
\global\belowdisplayskip = 18pt plus3pt minus9pt
\global\abovedisplayshortskip = 6pt plus3pt
\global\belowdisplayshortskip = 6pt plus3pt

\def\barsoff{\overfullrule=0pt}


\def\endignore{}
\def\ignore #1\endignore{} 

\newcount\dflag
\dflag = 0


\def\monthname{\ifcase\month 
\or January \or February \or March \or April \or May \or June%
\or July \or August \or September \or October \or November %
\or December 
\fi}

\newcount\dummy
\newcount\minute  
\newcount\hour
\newcount\localtime
\newcount\localday
\localtime = \time
\localday = \day

\def\advanceclock#1#2{ 
\dummy = #1
\multiply\dummy by 60
\advance\dummy by #2
\advance\localtime by \dummy
\ifnum\localtime > 1440 
\advance\localtime by -1440
\advance\localday by 1
\fi}

\def\settime{{\dummy = \localtime %
\divide\dummy by 60%
\hour = \dummy 
\minute = \localtime%
\multiply\dummy by 60%
\advance\minute by -\dummy 
\ifnum\minute < 10 
\xdef\spacer{0} 
\else \xdef\spacer{} 
\fi %
\ifnum\hour < 12 
\xdef\ampm{a.m.} 
\else 
\xdef\ampm{p.m.} 
\advance\hour by -12 %
\fi %
\ifnum\hour = 0 \hour = 12 \fi 
\xdef\timestring{\number\hour : \spacer \number\minute%
\thinspace \ampm}}}



\def\endtitle{}
\def\title#1\endtitle{\vskip.5in\titlefont
\global\baselineskip = 2\baselineskip 
#1\vskip.4in
\baselineskip = 0.5\baselineskip\rm}
 
\def\endauthors{}
\def\authors#1\endauthors{#1}

\def\endabstract{}
\def\abstract#1\endabstract{\vskip .3in%
\centerline{\sectionfont\bf Abstract}%
\vskip .1in
\noindent#1}

\def\nopageonenumber{\footline={\ifnum\pageno<2\hfil\else
\hss\tenrm\folio\hss\fi}}  

\newcount\nsection 
\newcount\nsubsection 

\def\section#1{\global\advance\nsection by 1
\nsubsection=0
\bigskip\noindent\centerline{\sectionfont \bf \number\nsection.\ #1}
\bigskip\rm\nobreak}

\def\subsection#1{\global\advance\nsubsection by 1
\bigskip\noindent\sectionfont \sl \number\nsection.\number\nsubsection)\
#1\bigskip\rm\nobreak}


\def\appendix#1#2{\bigskip\noindent%
\centerline{\sectionfont \bf Appendix #1.\ #2} 
\bigskip\rm\nobreak} 


\newcount\nref 
\global\nref = 1 

\def\therefs{} 


\def\ref#1#2{\xdef #1{[\number\nref]} 
\ifnum\nref = 1\global\xdef\therefs{\item{[\number\nref]} #2\ } 
\else
\global\xdef\oldrefs{\therefs}
\global\xdef\therefs{\oldrefs\vskip.1in\item{[\number\nref]} #2\ }%
\fi%
\global\advance\nref by 1
}

\def\listrefs{\vfill\eject\section{References}\therefs}


\newcount\nfoot 
\global\nfoot = 1 

\def\foot#1#2{\xdef #1{(\number\nfoot)} 
\hskip -0.2cm ${}^{\number\nfoot}$ 
\footnote{}{\vbox{\baselineskip=10pt
\eightrm \hskip -1cm ${}^{\number\nfoot}$ #2}}
\global\advance\nfoot by 1
}


\newcount\nfig 
\global\nfig = 1
\def\thefigs{} 

\def\figure#1#2{\xdef #1{(\number\nfig)}
\ifnum\nfig = 1\global\xdef\thefigs{\item{(\number\nfig)} #2\ }
\else
\global\xdef\oldfigs{\thefigs}
\global\xdef\thefigs{\oldfigs\vskip.1in\item{(\number\nfig)} #2\ }%
\fi%
\global\advance\nfig by 1 } 

\def\fig#1{\xdef #1{(\number\nfig)}
\global\advance\nfig by 1 } 


\newcount\ntab
\global\ntab = 1

\def\table#1{\xdef #1{\number\ntab}
\global\advance\ntab by 1 } 


\newcount\cflag
\newcount\nequation
\global\nequation = 1
\def\eqlabel{(1)}

\def\nexteqno{\ifnum\cflag = 0
\global\advance\nequation by 1
\fi
\global\cflag = 0
\xdef\eqlabel{(\number\nequation)}}

\def\lasteqno{\global\advance\nequation by -1
\xdef\eqlabel{(\number\nequation)}}

\def\label#1{\xdef #1{(\number\nequation)}
\ifnum\dflag = 1
{\escapechar = -1
\xdef\draftname{\littlefont\string#1}}
\fi}

\def\clabel#1#2{\xdef\eqlabel{(\number\nequation #2)}
\global\cflag = 1
\xdef #1{\eqlabel}
\ifnum\dflag = 1
{\escapechar = -1
\xdef\draftname{\string#1}}
\fi}

\def\cclabel#1#2{\xdef\eqlabel{#2)}
\global\cflag = 1
\xdef #1{\eqlabel}
\ifnum\dflag = 1
{\escapechar = -1
\xdef\draftname{\string#1}}
\fi}


\def\eeq{}

\def\eqnn #1\eeq{$$ #1 $$}

\def\eq #1\eeq{
\ifnum\dflag = 0
{\xdef\draftname{\ }}
\fi 
$$ #1
\eqno{\eqlabel \rlap{\ \draftname}} $$
\nexteqno}







\def\eqa #1\eeq{
\ifnum\dflag = 0
{\xdef\draftname{\ }}
\fi 
$$ \eqalignno{ #1 } $$
\global\cflag = 0}


\def\etal{{\it et.al.\/}}



\global\nulldelimiterspace = 0pt



\def\frac#1#2{{{#1} \over {#2}}\,}  



\def\Asl{\hbox{/\kern-.7500em\it A}} 
\def\Dsl{\hbox{/\kern-.6700em\it D}} 
\def\dsl{\hbox{/\kern-.5300em$\partial$}}
\def\pxpsl{\hbox{/\kern-.5600em$p$}}
\def\sslsh{\hbox{/\kern-.5300em$s$}}
\def\epssl{\hbox{/\kern-.5100em$\epsilon$}}
\def\delsl{\hbox{/\kern-.6300em$\nabla$}}
\def\lxpsl{\hbox{/\kern-.4300em$l$}}
\def\elxpsl{\hbox{/\kern-.4500em$\ell$}}
\def\kxpsl{\hbox{/\kern-.5100em$k$}}
\def\qxpsl{\hbox{/\kern-.5000em$q$}}
\def\sla#1{\raise.15ex\hbox{$/$}\kern-.57em #1}



\def\roughly#1{\mathrel{\raise.3ex\hbox{$#1$\kern-.75em\lower1ex\hbox{$\sim$}}}}







\def\pmb#1{\setbox0=\hbox{#1}%
\kern-.025em\copy0\kern-\wd0
\kern.05em\copy0\kern-\wd0
\kern-.025em\raise.0433em\box0}   


\font\jlgtenbrm=cmbx10
\font\jlgtenbit=cmmib10
\font\jlgtenbsy=cmbsy10
\font\jlgsevenbrm=cmbx10 at 7pt
\font\jlgsevenbsy=cmbsy10 at 7pt
\font\jlgsevenbit=cmmib10 at 7pt
\font\jlgfivebrm=cmbx10 at 5pt
\font\jlgfivebsy=cmbsy10 at 5pt
\font\jlgfivebit=cmmib10 at 5pt
\newfam\jlgbrm

\textfont\jlgbrm=\jlgtenbrm
\scriptfont\jlgbrm=\jlgsevenbrm
\scriptscriptfont\jlgbrm=\jlgfivebrm
\newfam\jlgbit

\textfont\jlgbit=\jlgtenbit
\scriptfont\jlgbit=\jlgsevenbit
\scriptscriptfont\jlgbit=\jlgfivebit
\newfam\jlgbsy

\textfont\jlgbsy=\jlgtenbsy
\scriptfont\jlgbsy=\jlgsevenbsy
\scriptscriptfont\jlgbsy=\jlgfivebsy
\newcount\jlgcode
\newcount\jlgfam
\newcount\jlgchar
\newcount\jlgtmp
\def\bolded#1{
        \jlgcode\the#1 \divide\jlgcode by 4096
        \jlgtmp\the\jlgcode \multiply\jlgtmp by 4096
        \jlgfam\the#1 \advance\jlgfam by -\the\jlgtmp
        \divide\jlgfam by 256
        \jlgtmp\the\jlgcode \multiply\jlgtmp by 16
	\advance\jlgtmp by \the\jlgfam
	\multiply\jlgtmp by 256
        \jlgchar\the#1 \advance\jlgchar by -\the\jlgtmp
        \advance\jlgfam by \the\jlgbrm
        \jlgtmp\the\jlgcode
        \multiply\jlgtmp by 16
        \advance\jlgtmp by \the\jlgfam
        \multiply\jlgtmp by 256
        \advance\jlgtmp by \the\jlgchar
        \mathchar\the\jlgtmp
}







\def\cm{{\rm \ cm}}

\input epsf.tex
\nopageonenumber
\baselineskip = 18pt
\barsoff

\def\bk{\item{}}

\title
\centerline{Footprints of the Newly-Discovered}
\vskip 0.1in
\centerline{Vela Supernova in Antarctic Ice Cores?}
\endtitle
\authors
\centerline{C.P. Burgess${}^a$ and K. Zuber${}^b$}  
\vskip .07in
\centerline{\it ${}^a$ Physics Department, McGill
University}
\centerline{\it 3600 University St., Montr\'eal, Qu\'ebec, 
CANADA, H3A 2T8.}
\vskip .07in
\centerline{\it ${}^b$ Lehrstuhl f\"ur Exp. Physik IV,
Universit\"at Dortmund}
\centerline{\it 44221 Dortmund, GERMANY.}
\endauthors
\vskip .25in
\abstract
The recently-discovered, nearby young supernova remnant in
the  southeast corner of the older Vela supernova remnant
may have been seen in measurements of nitrate abundances in
Antarctic ice cores. Such an interpretation of  this
twenty-year-old ice-core data would provide a more accurate
dating of this supernova than is possible purely using
astrophysical techniques. It permits an inference of the 
supernova4s ${}^{44}$Ti yield purely on an observational
basis, without reference to supernova modelling. The resulting estimates of the
supernova distance and light-arrival time are 200 pc and
700 years ago, implying an expansion speed  of 5,000 km/s
for the supernova remnant. Such an expansion speed has been
argued elsewhere to imply the explosion to have been a 15 
$M_\odot$ Type II supernova.  This interpretation also adds
new evidence to the debate as to whether nearby supernovae
can measurably affect nitrate abundances in polar ice cores.
\endabstract
\vfill\eject

\ref\BAsch{Bernd Aschenbach, {\it Discovery of a young
nearby supernova remnant}, {\sl Nature} v.~396, p.~141,
1998.}

\ref\Iyudin{A.F. Iyudin \etal, {\it Emission from
${}^{44}Ti$ associated with a previously unknown galactic
supernova},  {\sl Nature} v.~396, p.~142, 1998.}

\ref\Ashworth{W.B. Ashworth, {\it A probable Flamsteed
observation of the Cassiopeia supernova}, {\sl Jour. Hist.
Astron.} v.~11, p.~1, 1980.}

Only a handful of supernovae have exploded over the last
thousand years within several kpc of the Earth. To this
select group -- which is summarized \foot\cassA{The more
recent supernova Cassiopeia A  of around 1680 appears not to have
been widely seen, \hfil\break if it was seen at all ~\Ashworth.}  in
Table 1 -- there has recently been a new addition,  due to
the discovery of a young supernova remnant in ROSAT X-ray
data, RX J0852.0 - 4622, quite nearby \BAsch. This remnant
has RA $8^h \, 52^m$ and Declination $-46^o 22'$ (2000
epoch), and  in the likely event  that RX J0852.0 - 4622 is
identical to the COMPTEL Gamma Ray source GROJ0852-4642 it should be around
 200 pc away, with its light potentially first arriving at Earth as early as
 700 years ago  \Iyudin. 
\ref\Rood{Robert T. Rood, Craig L. Sarazin, Edward J.
Zeller and Bruce C. Parker, {\it $X-$ or $\gamma$-rays from
supernovae in glacial ice}, {\sl Nature}, v.~282, p.~701,
1979.}

Although there is no visual record of this supernova, its
proximity to the Earth suggests it might have left other
calling cards which might yet be found. To pursue this we
have searched the literature on geophysical supernova 
signatures. It is the purpose of this letter to point out that supernova RX
 J0852.0 - 4622 indeed appears to have left its 
mark, through its influence on the nitrate abundances in twenty-year-old ice
 cores which were drilled at the South Pole station. 

\midinsert
$$\vbox{\tabskip=0pt \offinterlineskip
\halign to \hsize{
\strut#& #\tabskip 1em plus 2em minus .5em&
#\hfil &#& \hfil#\hfil &#& \hfil#\hfil &#& \hfil#\hfil &#& 
\hfil#\hfil &#& \hfil#\hfil &#& \hfil#\hfil &#\tabskip=0pt\cr
\noalign{\hrule}\noalign{\smallskip}\noalign{\hrule}
\noalign{\medskip}
&& \hfil Date && Name && RA&& Dec && Visual
&& Distance &\cr  
&& && && (1950) && (1950) && Magnitude && (kpc) &\cr  
\noalign{\medskip}\noalign{\hrule}\noalign{\medskip}
&& 1006 3 April && --- && 15  10 && $-40$ && 
$-9.5$ && 1.3 &\cr  
&& 1054 4 July && Crab && 05  40 && $+20$ && 
$-4$ && 2.2 &\cr  
&& 1181 6 August && --- && 01  30 && $+65$ &&   
&& 2.6 &\cr  
&& 1572 8 November && Tycho && 00  20 && $+65$ 
&& $-4$ && 2.7 &\cr  
&& 1604 8 October && Kepler && 17  30 && $-20$ 
&& $-3$ && 4.2 &\cr  
\noalign{\medskip}\noalign{\hrule}\noalign{\smallskip}
\noalign{\hrule}
}}$$
\medskip
\centerline{\bf Table (1): Supernovae observations within
the last millenium.}   
\endinsert

In their original publication \Rood, the drillers of this ice core
identified within it three distinctive spikes in the nitrate abundance, whose
 dates of deposition correspond to the dates of the three latest supernova
 listed in Table I. (Their core sample was
not sufficiently deep to contain those of 1054 or 1006.) These
spikes are easily seen in Fig.~1, which is reproduced from  
Ref.~\Rood. Also seen in Fig.~1 is a fourth clear spike in the
nitrate abundance, which could not be attributed to any supernova
known at the time. It is remarkable that this fourth spike
corresponds precisely with the time when light -- including 
X- and gamma rays -- from the recently-discovered Vela supernova  would have
 been arriving at the Earth!

\ref\Risbo{T. Risbo, H.B. Clausen and K.L. Rasmussen,  {\it
Supernovae and nitrate in the Greenland ice sheet}, {\sl
Nature}, v.~294, p.~637, 1981.}

\ref\Parker{Bruce C. Parker, Edward J. Zeller and Anthony
J. Gow, {\it Nitrate fluctuations in Antarctic snow and
firn: potential sources and mechanisms of formation}, {\sl
Annals of Glaciology}, v.~3, p.~243, 1982.}

\ref\Herron{Michael Herron, {\it Impurity sources of $F^-$,
$Cl^-$, $NO_3^-$ and $SO_4^{2-}$ in greenland and Antarctic
precipitation}, {\sl Journal of Geophysical Research},
v.~87, no. C4, p.~3052, 1982.}

\ref\Damon{P.E. Damon, D. Kaimei, G.E. Kocharov, I.B. Mikheeva
and A.N. Peristykh, {\it Radiocarbon production by the gamma-ray
component of supernova explosions}, {\sl Radiocarbon}
{\bf 37} (1995) 599-604; 
P.E. Damon, G.E. Kocharov, A.N. Peristykh, I.B. Mikheeva
and K.M. Dai, {\it High energy gamma rays from SN1006AD},
{\sl Proc. 24th Int. Cosmic Ray Conf.}, Rome 1995, v. 2, p. 311-314.}

\ref\Ruderman{M.A. Ruderman, {\sl Science}, v.~184,
p.~1079, 1974.}

\ref\Whitten{R.C. Whitten, J. Cuzzi, W.J. Borucki and J.H.
Wolfe, {\it Effect of nearby supernova explosions on
atmospheric ozone}, {\sl Nature}, v.~263, p.~398, 1976.}

\ref\Clark{D.H. Clark, W.H. McCrea and F.R. Stephenson,
{\it Frequency of nearby supernovae and climatic and
biological  catastrophes}, {\sl Nature}, v.~265, p.~318,
1977.}

We have found no other geophysical signals for this supernova,
and our search for these unearthed an interesting controversy 
\Rood, \Risbo, \Parker, \Herron, regarding which the recent Vela supernova
 may shed new light.
The controversy concerns whether or not nearby supernovae 
can be detected by studying the concentration of nitrate
deposition  as a function of depth in polar ice cores.
Supernovae have been argued to have produced observable
changes in geophysical isotope abundances \Damon, and  
there is little question that supernovae can produce
$NO_3^-$ when the ionizing radiation they  generate
impinges on the atmosphere \Ruderman, \Whitten, 
\Clark.  What is not clear is whether this source of
atmospheric nitrates is detectable over other sources in
ice removed from polar core samples.

The evidence given in ref.~\Rood, that polar ice cores can register
$NO_3^-$ nitrate fluctuations of a cosmogenic origin, was
supported by observations of the nitrate abundance in 
Antarctic ice cores taken near the Vostok station (78${}^o$ 
28' S, 106 ${}^o$ 48' E) \Parker. The authors of ref.~\Parker\ 
claim to find evidence for a correlation between the nitrate
abundances and the cyclic variations in the 
solar activity. Because the overall nitrate deposition rate was 
found to be smaller in Vostok cores than in those from the 
South Pole, it was not possible to confirm at Vostok 
that nitrate abundances correlate with supernovae, although
(by eye) some increase in the nitrate levels is roughly 
coincident with the times of the various observed supernovae. 

The difference seen in the overall annual rate of nitrate 
fallout, which is lower at Vostok than at the South Pole \Parker,
might itself be some evidence in favour of its being of 
cosmogenic origin. As was observed in \Parker, such a
difference could arise if the nitrate production,
were associated with aurorae in the Antarctic
atmosphere.  Since aurorae occur when charged particles
impinge on the atmosphere, the geomagnetic field places 
them in a torus centred on the magnetic pole.  (Ionizing
bremstrahlung $X$-rays from these particles are also
directed downwards and so ionize the atmosphere
preferentially beneath the aurorae.) In the southern
hemisphere this makes aurorae more abundant over the South
Pole than over the Vostok station. If the nitrates
precipitate rapidly the nitrate abundance deposited on the
surface could also be higher at the South Pole than at
Vostok.  

\ref\TDchem{ 			
J.C. Stager, P.A. Mayewski, {\it Abrupt early to
mid-Holocene  		climate transition registered at the
equator and the poles}, 		{\sl Science} v. 276, p.
1834-1836.;\bk   
P.A. Mayewski, \etal, {\it Climate Change During the Last
Deglaciation in   Antarctica}, {\sl Science} v 272, p.
1636-1638 (1996);\bk  
E.J. Steig, \etal {\it Wisconsinan and Holocene climate
history from an   ice core at Taylor Dome, western Ross
Embayment, Antarctica}, 	{\sl Geografiska Annaler} (in
review).}

On the other hand, searches using Greenland ice cores in the early
1980's show no evidence for correlations between nitrate
levels and supernovae \Risbo, \Herron. We have ourselves
examined data for chemical depositions in ice cores taken
in the early 1990's from the Antarctic Taylor Dome
(77${}^o$ 48' S,  158${}^o$ 43' E --- reasonably close
to Vostok) \TDchem, and 
no spectacular nitrate peaking appears at depths corresponding 
to known supernovae (although some suggestive spikes do 
appear in the abundances of other ions, such as $Cl^-$).

The controversy emerges because these observations 
permit two different conclusions: 

\item{1.} 
Cosmogenic influences on
ion abundances in polar ice cores are swamped by terrestrial influences; or
\item{2.} 
Cosmogenic sources can detectably influence glacial ion abundances, 
but their fallout to the surface is uneven over the Earth's surface. 

To the supporters of option 1 the spikes of ref. \Rood\ must be 
due to some kind of experimental error, or to some
other kind of terrestrial source. Their correlation with
observed supernovae would be coincidental. For the supporters of 
option 2 the difficulty is understanding how cores taken at 
some places can carry cosmogenic signals, while those take at
others do not. 

Here we take the point of view that the agreement between
the newly-discovered supernova, RX J0852.0 - 4622,
and the fourth spike in the data of ref.~\Rood, makes 
coincidence a less convincing explanation for the remarkable
correlation between nitrate spikes and visible supernovae.
We therefore adopt the point of view of option 2,
in order to see what can be learnt about cosmogenic nitrate
deposition on the Earth, as well as about the properties of
the supernova itself.  We find that several inferences may
be drawn.  

\medskip\noindent {\it 1. The Smoking Gun:} 
First and foremost, the most obvious test of option 2 consists 
of further examination of ice cores taken
at the South Pole. Taking the spikes in the data of ref.~\Rood\
at face value means that some mechanism makes the nitrate 
fallout due to supernovae uneven around the globe, but has not 
removed the most recent signals at the South Pole itself. 
Although it is logically possible that the same mechanism might prevent
deeper South Pole cores from carrying the evidence of the earlier supernovae of
1054 and 1006, this possibility seems unlikely given the presence of
the four earlier spikes. Clearly, a comparison of the nitrate
levels in more ice cores -- especially those taken at the South Pole
which are deep enough to include these
last-mentioned  supernovae -- would be
very useful to clarify the experimental situation.
Furthermore one might envisage searching for signals in the
deposition rates of other chemical compunds like $Cl$ and
$NH_4$.

\ref\aschenbach{B.~Aschenbach, A.F.~Iyudin and V.~Sch\"onfelder,
{\it Constraints of age, distance and progenitor of the supernova
remnant  RX J0852.0 - 4622 / GRO J0852-4642}, {\sl Astronomy
and Astrophysics} (to appear), ({\tt astro-ph/9909415}).}
 
\medskip\noindent {\it 2. Dating the New SN Remnant:} 
The age ($t$) and distance ($d$) of RX J0852.0 - 4622 may be inferred
from the observed intensity ($f$) of the ${}^{44}$Ti decay gamma-ray line,
as well as the angular size ($\theta$) of the supernova remnant, using
\eq
\label\tthform
f = {1 \over 4 \pi d^2} \; \left( {Y_{44} \over m_{44}
\tau_{44}} \right) \; e^{-t/\tau_{44}}, \qquad\qquad
\theta = {v_m \;  t \over d}, 
\eeq
if the
${}^{44}$Ti yield ($Y_{44}$) of the supernova and the mean expansion velocity 
of its remnant ($v_m$) are known \aschenbach. (Here $\tau_{44} \approx 90$
yr is the ${}^{44}$Ti half life, and $m_{44}$ is its atomic mass.) $v_m$ 
may be inferred from the present velocity of
the shock wave, which is in turn found from the X-ray brightness which
it produces as it slams into the surrounding medium, and $Y_{44}$ is
taken from SN models. Not surprisingly, the values for $t$ and $d$ 
obtained in this way are subject to considerable uncertainty, with
age estimates being potentially inaccurate by hundreds of years. 

This chain of inference may be reversed if the nitrate spikes 
in the South Pole ice core are due to supernova RX J0852.0 - 4622,
because then more information is directly available from observations. 
For instance, a ${}^{44}$Ti yield of $5 \times 10^{-5} \; M_\odot$ may now
be directly inferred from observations given the date of the ice core spike 
(using the inferred mean remnant expansion speed of 5,000 km/s),
instead of being taken from numerical studies.  

Alternatively, profit may be made from the much better accuracy
with which the ages of the nitrate spikes in the South Pole 
ice core are known. In order to estimate the error
in determining the age for each depth of their South Pole
ice sample, Rood \etal\ provide three possible 
chronologies for the same core. These indicate the date of
the  previously-unidentified nitrate spike to be within the
range $1320 \pm 20$ AD.  If we take the ${}^{44}$Ti yield
from numerical models, then we may more precisely learn the expansion
velocity of the supernova ejecta. The agreement of 1320 AD with the
age determined from $X-$ray and $\gamma-$ray observations
of the supernova remnant then indicates that the ejecta
expansion velocity is close to the central value of 5,000 km/s
assumed in ref.~\Iyudin. 

Since the ratio between the intensity of two different gamma-ray
lines is independent of the distance to the SN remnant, more may
be learnt by comparing the intensity of the ${}^{44}$Ti line with
the ${}^{26}$Al line, which has also been observed. Using the half
lives $\tau_{44} \approx 90 \;\hbox{yr} \ll \tau_{26} \approx 1.07
\times 10^6$ yr, one finds in this way
\eq
{f_{44} \over f_{26}} = \left( {\tau_{26} \, m_{26} \, Y_{44} \over
\tau_{44} \, m_{44} \, Y_{26}} \right) \; e^{-t/\tau_{44}}.
\eeq
A complication arises in this case because 
although the short $\sim 90$ yr halflife of ${}^{44}$Ti ensures
the observed ${}^{44}$Ti gamma flux comes from  RX J0852.0 - 4622, 
the $1.07 \times 10^6$ 
yr half-life of ${}^{26}$Al makes it impossible to be sure that 
these gamma rays are not coming from the older Vela remnant rather
than just from RX J0852.0 - 4622. 

Two things may be learned here by assuming the ice-core date for
RX J0852.0 - 4622. First, if one uses the results 
of numerical models to infer an upper limit, 
$Y_{44}/Y_{26} < 100$ (or $<10$), together with the observed ${}^{44}$Ti flux,
$f_{44} = (3.8 \pm 0.7) \times 10^{-5}$/cm${}^2$/s, 
then one finds a lower limit to the ${}^{26}$Al
flux from  RX J0852.0 - 4622: $f_{26} > 1 \times 10^{-7}$/cm${}^2$/s
(or $ > 1 \times 10^{-6}$/cm${}^2$/s). 
Alternatively, if the observed point-source flux, $f_{\rm pt} =  
(2.2 \pm 0.5) \times 10^{-5}$/cm${}^2$/s, of ${}^{26}$Al gamma rays
is assumed to be coming from  RX J0852.0 - 4622, then we learn
$Y_{44}/Y_{26} \approx 0.5$ (which is in agreement with ref.~\aschenbach\
provided $v_m = 5,000$ km/s).

\ref\Chen{W. Chen and N. Gehrels, {\it The progenitor of
the new  COMPTEL/ROSAT supernova remnant in Vela}, {\sl
Astrophysical Journal Letters} v. 514, p. L103 (1999),
({\tt astro-ph/9812154 v2}).}

\ref\Halpern{J.P. Halpern and S.S. Holt, {\it Discovery of
soft $X-$ray pulsations from the $\gamma-$ray source
Geminga}, {\sl Nature}, v.~357, p.~222, 1992.}
\ref\Bignami{G.F. Bignami, P.A. Caraveo and S. Mereghetti,
{\it The proper motion of Geminga's optical counterpart},
{\sl Nature}, v.~361, p.~704, 1993.}
\ref\Wang{Daniel Wang, Zhi-Yun Li and Mitchell C. Begelman,
{\it The $X-$ray emitting trail of the nearby pulsar
PSR1929+10}, {\sl Nature}, v.~364, p.~127, 1993.}

\ref\Castagnoli{G. Cini Castagnoli and G. Bonino,  {\it
Thermoluminescence in sediments and historical supernovae
explosions}, {\sl Il Nuovo Cimento} v.~5C, n.~4, p.~488,
1982.}

\ref\Ellis{John Ellis, Brian D. Fields and David N.
Schramm, {\it Geological isotope anomalies as signatures of
nearby supernovae}, ({\tt astro-ph/9605128}).}

\ref\Fields{Brian D. Fields and John Ellis,  {\it On deep
ocean ${}^{60}Fe$ as a fossil of a near-earth supernova},
preprint CERN-TH/98-373 ({\tt astro-ph/9811457}).}

\medskip\noindent
{\it 3. The Nature of the Supernova Explosion:} 

As is argued in Ref.~\Chen, an expansion velocity this
large for the SNR argues that this was a 15 $M_\odot$ Type
II supernova. Moreover an age of 700 years gives further
information, as can be seen from Fig.~2. Having a
relatively nearby Vela supernova of less than 250 pc, gives
a ${}^{44}$Ti yield of less than $10^{-4} M_{\odot}$. This
disfavours a Type Ia supernova explosion within a dense
region as a possible progenitor of the new Vela supernova. 

Since it was so nearby, it would be worthwhile to look for
other cosmogenic signals for this supernova, such as have
been proposed for the very nearby Geminga event several
hundred thousand years ago \Halpern, \Bignami, \Wang,
through enhancements in the abundances of radionuclides
in  sediments \Castagnoli, \Ellis, 
\Fields.  One might imagine even searching for signals due
to the neutrino flux, since this  should be as large as
$10^{16}\cm^{-2}{\rm s}^{-1}$. 

\medskip\noindent
{\it 4. The Distance to SN1006:} 
As was already noticed in \Iyudin, this date for
the arrival  time of light from the supernova implies the
supernova distance must be 200 pc, which is on the near
side of the range which is allowed by the $X-$ray
measurements. This makes this the closest supernovae which
happened in the last millenium. Since this range was
determined \BAsch\ by comparing the brightness of remnant 
RX J0852.0 - 4622 with the remnant of the 1006 supernova,
the 1006 remnant  must be about 800 pc away, which is also
at the near end of its allowed range. 

\medskip\noindent {\it 5. The Distance-Dependence of the Nitrate Signal:}
It is tempting to observe that a distance of 200 pc to  
RX J0852.0 - 4622 makes this supernova 10 times closer than 
the next nearest SN remnant listed in Table 1. This raises the question
as to why the flux of ionizing radiation was not therefore 100 times 
as large for this supernova than for all of the others, with 
a correspondingly large nitrate peak. Such a large variation in 
amplitude is clearly not visible for the peaks in Fig.~(1). 

We have three reasons not to be disturbed by this naive 
factor of 100 in radiation intensity. First, as mentioned in
the previous item, the distance estimates to the supernovae of
Table (1) carry relatively large uncertainties, with the remnant 
of SN1006 being possibly only 4 times as distant as RX J0852.0 - 4622.
Second, all supernovae are not alike and two supernovae 
can differ widely in their brightness even if they are 
equidistant. (This point is perhaps most dramatically illustrated by
the nonobservation of the supernova associated with the 
Cass A remnant.) Third, given that some poorly-understood mechanism
is required to ensure that the rate of nitrate fallout is not uniform
around the globe --- as must be assumed if we are to interpret
the Rood \etal\ spikes as being cosmogenic in origin --- we should
expect no simple connection between the size of a nitrate spike and
the amount of ionizing radiation received at the Earth. 

\medskip\noindent {\it 6. The Distribution of Nitrate Fallout:} 
Finally, if the four nitrate spikes of the Rood \etal\ core 
are really associated with supernovae, then it still
must be understood why nitrate levels of supernova origin
are unevenly deposited around the globe, and why they are
larger at the South Pole than they are in Greenland and 
elsewhere in Antarctica. 

\ref\Crutzen{P.J. Crutzen, {\it Ozone production rates in
an      oxygen-hydrogen-nitrogen oxide atmosphere}, {\sl
Journal    of Geophysical Research}, v 30, p. 7311-7327
(1976).}

One possibility is suggested if the ionization mechanism
due to the supernova were associated with aurorae. Besides
potentially explaining different nitrate deposition rates
at different Antarctic sites if the settling rate is
sufficiently fast, aurorae might also account  for
differences between the northern and southern hemispheres.
For auroral production produced by protons directed to the
Earth by solar flares, the conversion to $NO_3^-$ proceeds
mainly at night \Crutzen, and so at high latitudes nitrate
production proceeds most abundantly during the winter.
Since the five supernovae listed in Table (1) all occur
between April and early October, nitrate deposition in
the  northern hemisphere could be less efficient if the
connection  between supernovae and aurora were also to
cause more effective  nitrate production during the
southern winter. 

Of course there are also several problems with this kind of
mechanism, which would have to be understood. First, association
with aurorae usually means the ionization is accomplished by
charged particles which preferentially hit the atmosphere near
the magnetic poles because they move along the magnetic field lines
of the Earth. But charged particles are not likely to have 
reached us yet from RX J0852.0 - 4622, since cosmic rays
diffuse through the interstellar medium and would take tens
of thousands of years to travel the intervening 200 pc.
In addition, any such aurora-based scenario must also
explain the absence of a solar-cycle dependence in the
deposition rate in cores taken near the north magnetic pole.

It is our hope that the remarkable correspondence 
between the arrival
time of light from RX J0852.0 - 4622, and the date of 
ref.~\Rood's fourth spike will stimulate further progress in
understanding the nature of terrestrial signals for nearby violent
astrophysical events. 

\bigskip
\centerline{\bf Acknowledgments}
\bigskip

This research was partially funded by N.S.E.R.C.\ of Canada
and les   Fonds F.C.A.R.\ du Qu\'ebec. We thank John Beacom
for updating us on the supernova distances listed in Table 1.

\bigskip
\centerline{\bf Figure captions}
Fig.~1: Orginal data on nitrate abundance as obtained by Rood 
\etal\ for a South Pole ice core.
Clearly visible are the spikes which can be associated to
supernova explosions.

\ref\Wietfeldt{F.E. Wietfeldt et al., {\it Long-term
measurement of the     half-life of $^{44}$Ti}, {\sl Phys.
Rev. C},      v 59, p. 528-530 (1999).}
\ref\Norman{E.B. Norman et al., {\it Half-life of
$^{44}$Ti},      {\sl Phys. Rev. C}, v 57, p. 2010-2016
(1998).}

Fig.~2: Distance versus ${}^{44}$Ti yield for two assumed
lifetimes of  ${}^{44}$Ti for a given supernova 700 years
ago. The distance $d$ is determined by the gamma flux $f_{44}$
and ${}^{44}$Ti - lifetime using the quadratic distance dependence
of the ejected ${}^{44}$Ti mass $Y_{44} = 4\pi e^{t/\tau} m_{44} \tau f_{44} d^2$.
The lifetimes
used are 87.5 years (dashed line)
\Wietfeldt{} and 90.4 years (solid line) \Norman. As can
be seen, for reasonable distances (below 250 pc) to the new
Vela supernova remnant the ${}^{44}$Ti yield is always
below $10^{-4}M_{\odot}$. This disfavours SN Ia explosions
in dense regions as a possible progenitor of the new Vela
supernova.

\listrefs

\vfill\eject
\centerline{\epsfxsize=11.5cm\epsfbox[45 430 550 750]{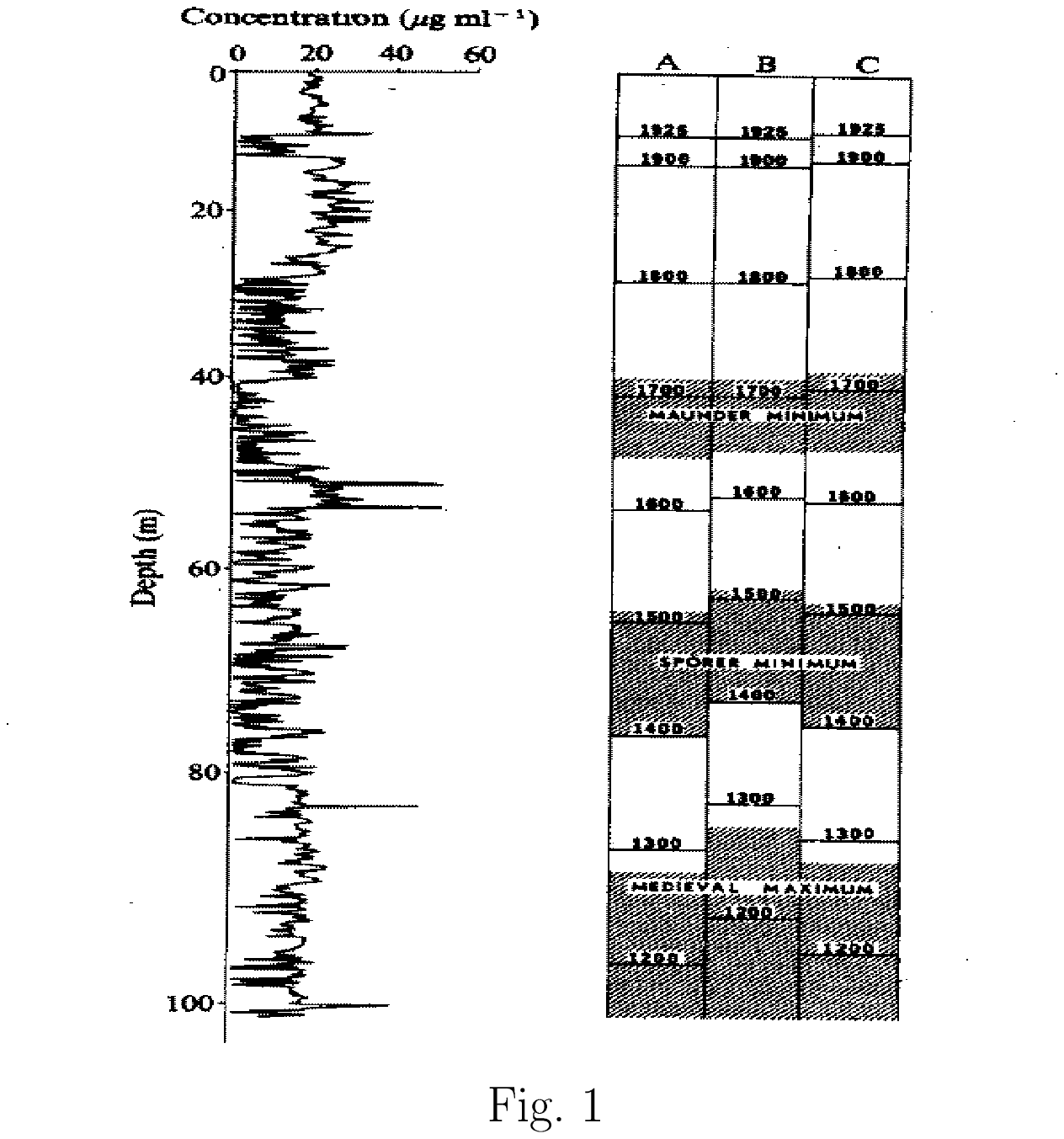}}

\vfill\eject
\centerline{\epsfxsize=11.5cm\epsfbox[45 430 550 750]{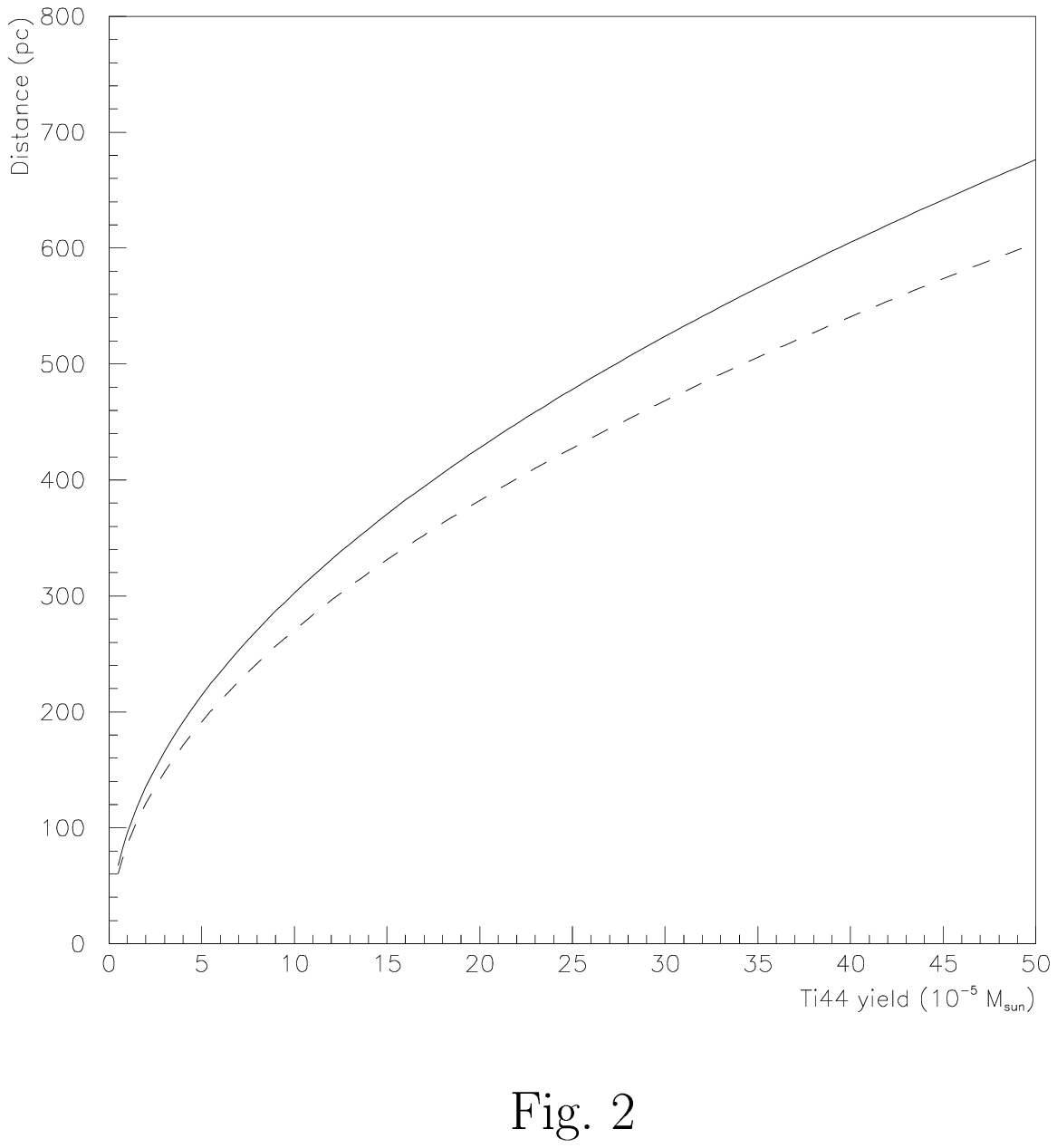}}
\bye